\documentclass[twocolumn,prl,amsmath,amssymb,superscriptaddress]{revtex4-1}
\usepackage[utf8]{inputenc}
\usepackage{amsfonts,amsmath,amssymb}
\usepackage{graphicx}
\usepackage{bm} 
\usepackage{mathrsfs}
\usepackage{subfigure}
\usepackage{textcomp}
\usepackage{hyperref}
\usepackage[flushleft]{threeparttable}
\usepackage[per-mode=symbol]{siunitx}
\usepackage[version=3]{mhchem}

\graphicspath{{../figures/}}

\DeclareSIUnit\intensity{\watt\per\centi\meter\squared}
\DeclareSIUnit\fieldstrength{\volt\per\centi\meter}
\usepackage{xspace}

\DeclareUnicodeCharacter{2009}{\,}

\newcommand{\cost}{\ensuremath{\langle\cos^2\theta_\text{2D}\rangle}}

\newlength{\figwidth}
\newlength{\figwidthwide}
\let\orgautoref\autoref
\providecommand{\Autoref}{%
  \def\equationautorefname{Equation}%
  \def\figureautorefname{Figure}%
  \def\subfigureautorefname{Figure}%
  \orgautoref}
\renewcommand{\autoref}{%
  \def\equationautorefname{Eq.}%
  \def\figureautorefname{Fig.}%
  \def\subfigureautorefname{Fig.}%
  \orgautoref}

\usepackage{color}
\usepackage[normalem]{ulem}
\definecolor{darkgreen}{rgb}{0.0,0.7,0.0}

\begin{document}


\title{Nonadiabatic laser-induced alignment dynamics of molecules on a surface} 



\author{Lorenz Kranabetter}
\affiliation{Department of Chemistry, Aarhus University, Langelandsgade 140, DK-8000 Aarhus C, Denmark}

\author{Henrik H. Kristensen}
\affiliation{Department of Physics and Astronomy, Aarhus University, Ny Munkegade 120, DK-8000 Aarhus C, Denmark}

\author{Areg Ghazaryan}
\affiliation{Institute of Science and Technology Austria, Am Campus 1, 3400 Klosterneuburg, Austria}

\author{Constant A. Schouder}
\affiliation{Department of Chemistry, Aarhus University, Langelandsgade 140, DK-8000 Aarhus C, Denmark}
\affiliation{Universit\'{e} Paris-Saclay, CEA, CNRS, LIDYL, 91191 Gif-sur-Yvette, France}

\author{Adam S. Chatterley}
\affiliation{Department of Physics and Astronomy, Aarhus University, Ny Munkegade 120, DK-8000 Aarhus C, Denmark}

\author{Paul Janssen}
\affiliation{Anorganisch-Chemisches Institut, Ruprecht-Karls-Universität Heidelberg, Im Neuenheimer Feld 270, 69120 Heidelberg, Germany}


\author{Frank Jensen}
\affiliation{Department of Chemistry, Aarhus University, Langelandsgade 140, DK-8000 Aarhus C, Denmark}

\author{Robert E. Zillich}
\affiliation{Institute for Theoretical Physics, Johannes Kepler Universit\"{a}t Linz, Altenbergerstraße 69, A-4040 Linz, Austria}

\author{Mikhail Lemeshko}
\affiliation{Institute of Science and Technology Austria, Am Campus 1, 3400 Klosterneuburg, Austria}

\author{Henrik Stapelfeldt}
\email[]{henriks@chem.au.dk}
\affiliation{Department of Chemistry, Aarhus University, Langelandsgade 140, DK-8000 Aarhus C, Denmark}

\date{\today}

\begin{abstract}

We demonstrate that a sodium dimer, \ce{Na2}(1~$^3\Sigma_{u}^+$), residing on the surface of a helium nanodroplet, can be set into rotation by a nonresonant 1.0 ps infrared laser pulse. The time-dependent degree of alignment measured, exhibits a periodic, gradually decreasing structure that deviates qualitatively from that expected for gas phase dimers. Comparison to alignment dynamics calculated from the time-dependent rotational Schr\"{o}dinger equation shows that the deviation is due to the alignment dependent interaction between the dimer and the droplet surface. This interaction confines the dimer to the tangential plane of the droplet surface at the point where it resides and is the reason that the observed alignment dynamics is also well-described by a 2D quantum rotor model.

\end{abstract}


\maketitle 

Through the nonresonant polarizability interaction, moderately intense femtosecond or picosecond laser pulses can create rotational wave packets, i.e. coherent superpositions of rotational eigenstates, in gas-phase molecules. Such wave packets are the foundation for laser-induced nonadiabatic alignment dynamics, where molecules exhibit alignment and anti-alignment in narrow, periodically occurring time windows, termed revivals. Nonadiabatic alignment has been explored, developed and applied in a huge number of works over the past 25 years~\cite{stapelfeldt_colloquium:_2003,ohshima_coherent_2010,fleischer_molecular_2012,koch_quantum_2019}. Rotational wave packets are also the basis for the even older discipline of rotational coherence spectroscopy (RCS) where rotational constants of molecules are obtained by comparing simulated and measured time-dependent alignment-sensitive experimental observables~\cite{felker_rotational_1992,riehn_high-resolution_2002}.


Recently, it was shown that nonadiabatic alignment~\cite{pentlehner_impulsive_2013,shepperson_laser-induced_2017} and RCS~\cite{chatterley_rotational_2020,cherepanov_excited_2021,qiang_femtosecond_2022} also applies to molecules solvated in liquid helium, in practice implemented by embedding molecules in nanometer-sized droplets of superfluid helium~\cite{toennies_superfluid_2004,choi_infrared_2006}. At comparatively low laser intensities, the measured alignment dynamics of \ce{OCS}, \ce{CS2} and \ce{I2} molecules was well reproduced by dynamics calculated from the time-dependent rotational Schr\"{o}dinger equation of gas-phase molecules subject to the polarizability interaction with the laser pulse, taking into account the effective rotational and centrifugal distortion constants of the in-droplet molecules~\cite{chatterley_rotational_2020}. Here, we study femtosecond-laser-induced alignment of molecules in a regime that to our knowledge is unexplored, namely on a surface. The studies concern alkali dimers, exemplified by \ce{Na2}, which are formed and residing on the surface of He nanodroplets~\cite{stienkemeier_use_1995,higgins_helium_1998, toennies_superfluid_2004}. Our motivation for the work is twofold. Firstly, we want to explore if nonresonant femtosecond or picosecond laser pulses can induce characteristic alignment dynamics and produce a noticeable degree of alignment for molecules on a surface and if so, does the potential binding the dimers to the surface change the alignment dynamics compared to that of molecules in the gas phase or inside He droplets? Secondly, although frequency-resolved UV/VIS spectroscopy and vibrational wave packet techniques have been applied extensively to investigate the electronic and vibrational states of alkali dimers on He droplets~\cite{stienkemeier_laser_1995, bruhl_triplet_2001, mudrich_formation_2004, claas_wave_2006, aubock_triplet_2007, gruner_vibrational_2011, lackner_spectroscopy_2013, sieg_desorption_2016}, no information has been obtained about their rotational energy levels. Is it possible that the RCS aspect of nonadiabatic alignment dynamics can provide useful insight here?

The experiment is conducted with the setup described in~\cite{kristensen_laser-induced_2023}. In brief, a continuous beam of He droplets, estimated to contain on average 15000 He atoms, is sent through a pick-up cell containing a gas of \ce{Na} atoms. The vapor pressure is adjusted such that some of the droplets pick up two \ce{Na} atoms, which leads to the formation of a sodium dimer, \ce{Na2,} in either the 1~$^1\Sigma_{g}^+$ ground state or in the lowest-lying triplet state 1~$^3\Sigma_{u}^+$. Then the droplet beam enters a velocity map imaging (VMI) spectrometer, in the center of which it is crossed by two focused, linearly polarized laser beams. The pulses in the first beam ($\lambda = 1.30~\mu$m, $\tau_\text{FWHM} = 1.0$~ps, $I_0 = \SI{5.0e10}{\intensity}$) induce alignment of the dimers. The wavelength of 1.30 $\mu$m is chosen to ensure nonresonant conditions, i.e. the photon energy is not resonant with electronic transitions in the dimers. The probe pulses in the second beam ($\lambda = 800$~nm, $\tau_\text{FWHM} = 50$ fs, $I_0 = \SI{5.0e12}{\intensity})$, are used to Coulomb explode the dimers into a pair of \ce{Na+} ions~\cite{kristensen_quantum-state-sensitive_2022,kristensen_laser-induced_2023}. From the emission directions of the \ce{Na+} fragment ions, recorded by a 2D imaging detector at the end of the VMI spectrometer, the degree of alignment of the dimers at time $t$ is determined, $t$ being the delay between an alignment and a probe pulse.


\begin{figure}
\includegraphics[width=8.5 cm]{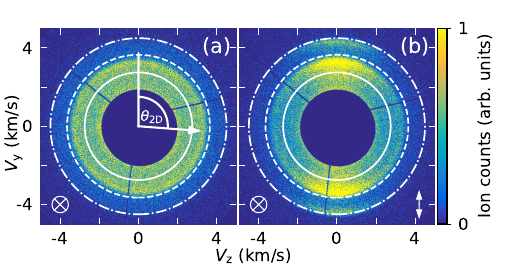}%
\caption{2D-velocity image of \ce{Na+} ions for (a) probe pulse only and (b) alignment and probe pulse, $t$ = 6.0 ps. The polarization directions of the probe ($\otimes$) and aligment ($\updownarrow$) laser pulses are indicated below the images. The three annotated white circles mark the two channels for ions from Coulomb explosion of dimers in the 1~$^1\Sigma_{g}^+$ state (outer channel) and in the 1~$^3\Sigma_{u}^+$ state (inner channel), see text.}
\label{fig:images}
\end{figure}

\Autoref{fig:images} shows 2D velocity images of \ce{Na+} ions obtained for He droplets doped with sodium. The image in \autoref{fig:images}(a) is recorded with the probe pulse only whereas for the image in \autoref{fig:images}(b) the alignment pulse was included. The central circular area and the three radial stripes void of signal are due to a metal disk and its supports installed in front of the imaging detector~\cite{chatterley_laser-induced_2020}. The disk blocks the large number of low-velocity \ce{Na+} ions coming from ionization of droplets doped with a single \ce{Na} atom or from ionization of isolated Na atoms that diffused into the VMI spectrometer, and partly the \ce{Na+} ions from dissociative ionization of dimers~\cite{kristensen_quantum-state-sensitive_2022,kristensen_laser-induced_2023}. Two distinct, radially separated channels are visible, their boundaries are marked by white annotated circles. The \ce{Na+} ions between the annotated solid and dashed (dashed and dot-dashed) circles originate from double ionization of \ce{Na2} in the 1~$^3\Sigma_{u}^+$ (1~$^1\Sigma_{g}^+$) state and Coulomb explosion of \ce{Na2^{2+}} into a pair of \ce{Na+} ions~\cite{kristensen_quantum-state-sensitive_2022,kristensen_laser-induced_2023}. An energy diagram of the potential curves for the two \ce{Na2} states and for \ce{Na2^{2+}} is given in the Supplemental Material. Thus, by separately analyzing the ion hits in the two channels, we can characterize the alignment dynamics of the dimers in each of the two different quantum states. In this work we focus on the \ce{Na2}(1~$^3\Sigma_{u}^+$) results and report the results for \ce{Na2} (1~$^1\Sigma_{g}^+$) elsewhere~\cite{kristensen-preparation}.

The image with the probe pulse only, \autoref{fig:images}(a), is circularly symmetric consistent with the expectations of randomly oriented dimers. By contrast, in the image recorded 6.0 ps after the alignment pulse, \autoref{fig:images}(b), the ions in both the innermost and the outermost channel localize along the alignment pulse polarization (vertical). This shows that at $t$ = 6.0 ps, dimers in both the 1~$^3\Sigma_{u}^+$ state and in the 1~$^1\Sigma_{g}^+$ state are aligned. To measure the alignment dynamics, we recorded images of \ce{Na+} ions from $t = -20$~ps to $t = 1200$~ps in steps of 1 ps. For each image, the degree of alignment $\cost$ was determined from the ion hits in the 1~$^3\Sigma_{u}^+$ state radial range. Here $\theta_\text{2D}$ is the angle in the detector plane between the ion hit and the alignment pulse polarization, see \autoref{fig:images}(a).

\begin{figure}
\includegraphics[width=8.5 cm]{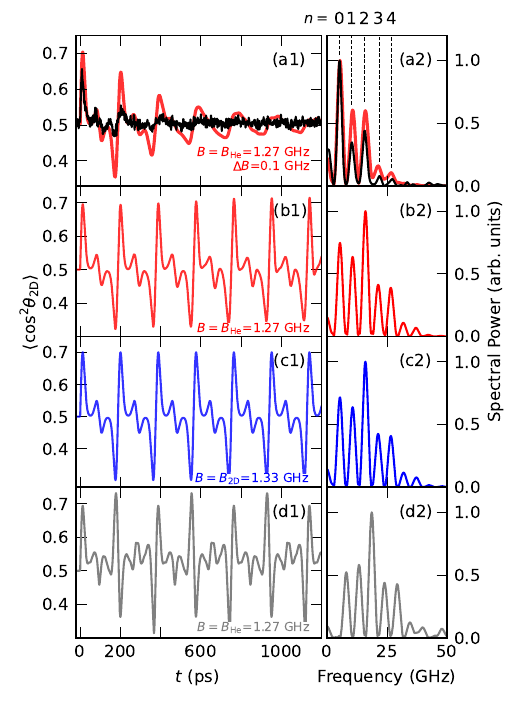}%
\caption{Left column: Time-dependent degree of alignment for \ce{Na2}(1~$^3\Sigma_{u}^+$). (a1): Experimental results (black curve) and simulated results (red curves) obtained with the 3D + $V_\text{He}$ model where inhomogeneous broadening is included. (b1): Simulation, 3D + $V_\text{He}$ model. (c1): Simulation, 2D model (essentially identical to the outcome of the 3D + $V_\text{He}$ model). (d1): Simulation, 3D model. The value of $B$ used in the different calculations is given in each panel. Right column: The power spectra of the corresponding $\cost$ traces. The spectral peaks are numbered according to their order, $n$, given on top of the panels.}
\label{fig-traces}
\end{figure}

The black curve in \autoref{fig-traces}(a1) depicts \cost($t$). It rises from the initial value of $\sim$~0.5 to a maximum of 0.66 at $t$ = 6 ps, demonstrating alignment after the pulse is turned off i.e.\ under field-free conditions. At longer times, \cost($t$) exhibits an oscillatory structure with recurring maxima and minima that gradually decrease in amplitude and broaden. This time-dependence of $\cost$ deviates qualitatively from previous results for linear molecules in the gas phase~\cite{dooley_direct_2003,wu_nonadiabatic_2011,thomas_hyperfine-structure-induced_2018} or inside He droplets~\cite{chatterley_rotational_2020}.

\begin{figure}
\includegraphics[width=8.5 cm]{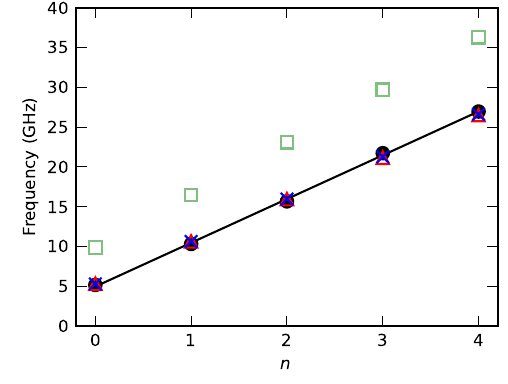}%
\caption{Central frequencies of the peaks in the power spectra versus $n$. Black circles: Experimental results, \autoref{fig-traces}(a2). Red triangles: 3D + $V_\text{He}$ model, \autoref{fig-traces}(b2). Blue crosses: Simulated results, 2D model, \autoref{fig-traces}(c2). The full black line represents the best linear fit to the experimental points. Green squares: Gas phase frequencies, (4$J$~+~6$)B_\text{gas}$, with $J=n$.  }
\label{fig-spectral-positions}
\end{figure}

To analyze the spectral content of the alignment trace in \autoref{fig-traces}(a1), \cost($t$) was Fourier transformed. The power spectrum shown in \autoref{fig-traces}(a2), contains discrete peaks numbered by $n$ = 0,1,2,3,4. The central positions of these spectral peaks plotted as a function of $n$, filled black circles in \autoref{fig-spectral-positions}, fall almost exactly on a straight line and the best linear fit is represented by the black line. In comparison, for linear molecules in the gas phase, the peak positions are given by (4$J$~+~6$)B_\text{gas}$ (ignoring centrifugal distortion~\cite{chatterley_rotational_2020}), where $B_\text{gas}$ is the rotational constant, equal to 1.65 GHz for \ce{Na2}(1~$^3\Sigma_{u}^+$)~\cite{bauer_accurate_2019} and $J$ the rotational quantum number. These peak positions are represented by the green squares in \autoref{fig-spectral-positions} and they differ strongly from the experimental results. Thus, the spectral analysis also reveals the deviation of the measured alignment dynamics from gas phase behavior.

To interpret the experimental observations, we calculated \cost($t$) by solving the time-dependent rotational Schr\"{o}dinger equation with the following Hamiltonian:
\begin{align}
 \hat{H}=B\hat{J}^2 - \frac{1}{4}E(t)^2  \Delta\alpha \cos^2(\theta)  + \hat{V}_\text{He}(\Theta).\label{eqn:2a}
\end{align}
The two first terms are those used to describe laser-induced alignment of isolated, linear, gas-phase molecules~\cite{friedrich_alignment_1995}. Here, $E(t)$ is the electric field envelope of the alignment pulse and $\theta$ the angle between the internuclear axis and the polarization of the alignment pulse. Furthermore, $\hat{J}^2$ is the squared rotational angular momentum operator, $B$ the rotational constant and $\Delta\alpha$ the polarizability anisotropy. We assume that $\Delta\alpha$ is given by the value of an isolated dimer~\cite{deiglmayr_calculations_2008} while for $B$ we employ an effective rotational constant $B_\text{He}$ similar to that used in the characterization of rotational states of molecules inside He droplets~\cite{choi_infrared_2006}. In the calculations, $B_\text{He}$ is treated as a free parameter as in RCS~\cite{riehn_high-resolution_2002}, and its value is varied to optimize the agreement between the calculated and measured $\cost$. The third term is an effective mean-field potential generated by the interaction between \ce{Na2} and the He droplet~\cite{guillon_theoretical_2011} with $\Theta$ denoting the angle between the internuclear axis and the droplet surface normal. Using path integral Monte Carlo (PIMC) simulations we calculated the angular distribution, $p(\Theta)$, for the internuclear axis~\cite{guillon_theoretical_2011}. From $p(\Theta)$ we then retrieved $V_\text{He}$ by approximating the pendular motion of \ce{Na2} as an effective one-body problem. See Supplemental Material for details. \Autoref{fig-potentials-probability} shows $p(\Theta)$ and $V_\text{He}$. Similar to \ce{Li2}~\cite{bovino_spin-driven_2009} and \ce{Rb2}~\cite{guillon_theoretical_2011}, $p(\Theta)$ peaks for $\Theta=\pi/2$, i.e.\ \ce{Na2} is preferably oriented parallel to the surface, although the distribution is wider than in the \ce{Rb2} case \cite{guillon_theoretical_2011}. The dimple in the $^4$He surface made by \ce{Na2} is depicted in the inset, shown as the He density isosurface at half its equilibrium density. Compared to \ce{Rb2}, \ce{Na2} leaves a deeper impression in the surface.

\begin{figure}
\includegraphics[width=\columnwidth]{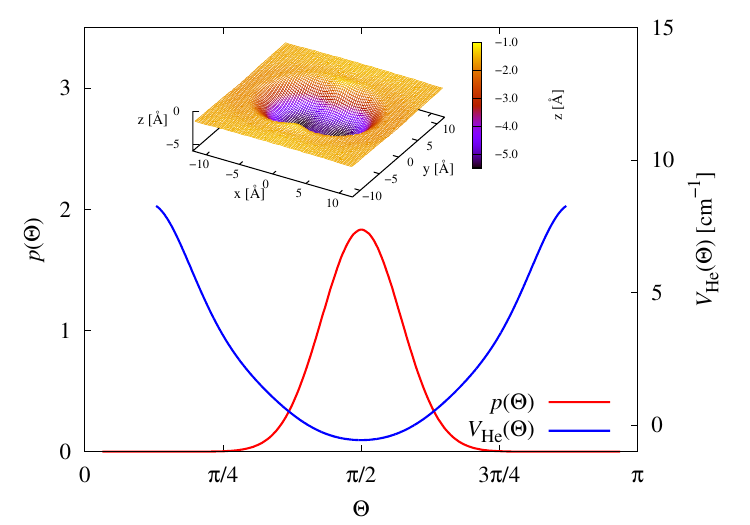}%
\caption{Angular distribution function $p(\Theta)$ (red) and the effective potential between \ce{Na2}(1~$^3\Sigma_{u}^+$) and the helium surface $V_\text{He}(\Theta)$ (blue), derived from $p(\Theta)$. $\Theta$ is the angle between the internuclear axis of the molecule and the surface normal. The inset shows the dimple in the He surface made by \ce{Na2}(1~$^3\Sigma_{u}^+$).}
\label{fig-potentials-probability}
\end{figure}

\Autoref{fig-traces}(b1) shows the calculated $\cost$ for $B_\text{He}$ = 1.27 GHz. The calculations were averaged over the initially populated rotational states, assuming a Boltzmann distribution with $T$ = 0.37 K, and over the focal volume determined by the measured spot sizes of the alignment and probe laser beams. The positions of the maxima and minima in $\cost$ match those of the experiment. Also, the positions of the five peaks in the calculated spectrum, \autoref{fig-traces}(b2), agree with the experimental findings.
However, the calculated $\cost$ does not reproduce the gradual decrease and broadening present in the experimental trace nor do the relative amplitudes of the spectral peaks match those in the experimental spectrum.

Recently, it was shown that for molecules inside He droplets, inhomogeneous broadening of the rotational levels must be included in the theoretical model to reproduce the experimental $\cost$ traces~\cite{chatterley_rotational_2020}. As it seems plausible that rotation of alkali dimers on He droplet surfaces are also subject to inhomogeneous broadening due to different droplet sizes~\cite{lehmann_potential_1999,lehmann_lorentzian_2007,zillich_lineshape_2008}, we implemented this effect through a Gaussian distribution of  $B_\text{He}$ with a FWHM, $\Delta B_\text{He}$~=~0.1~GHz~\footnote{This gives approximately the same $\Delta B$/$B$ ratio as for \ce{CS2} and \ce{I2} in \cite{chatterley_rotational_2020}}. The resulting $\cost$($t$), represented by the red curve in \autoref{fig-traces}(a1) shows gradual decay and broadening of the maxima and minima similar to the experimental trace. In addition, the relative amplitudes of the peaks in the spectrum, \autoref{fig-traces}(a2), are now much closer to those of the experimental peaks. The experimentally observed decrease of \cost($t$) occurs, however, faster than that predicted by the simulations indicating that the rotational levels are also subject to lifetime (homogeneous) broadening where the lifetime of the rotational states decreases when the rotational energy increases. The current experimental scheme does not allow a distinction between the different broadening mechanisms. That should, however, be possible by e.g.\ rotational echo techniques~\cite{karras_orientation_2015,rosenberg_echo_2018,zhang_rotational_2019}, which could open opportunities for time-resolved probing of the coupling of the rotating dimer to the surface or bulk modes of the He droplet~\cite{krotscheck_dynamics_2001}. Also, recording alignment dynamics at different droplet sizes may provide insight into the broadening mechanism.


Regarding the RCS aspect of nonadiabatic alignment dynamics, we note that for linear molecules in the gas phase, the spectral lines represent the coupling of two stationary rotational states with quantum numbers $J$ and $J+2$ or $J-2$~\cite{sondergaard_nonadiabatic_2017}. This gives the characteristic central positions of the spectral lines, $(4J+6)B_\text{gas}$, which makes it straightforward to determine $B_\text{gas}$~\footnote{and also $D_\text{gas}$ if centrifugal distortion is included in the analysis} from experimental spectra~\cite{sussman_quantum_2006,schroter_crasy:_2011,przystawik_generation_2012,zhang_time-domain_2018}. For alkali dimers on the surface of He droplets, the dimer-surface interaction implies that the eigenstates are no longer pure free rotor states. This will lead to a deviation from the $(4J+6)B$ expression. The experiment and the simulation showed that in the case of \ce{Na2}(1~$^3\Sigma_{u}^+$), there is still a linear relation between the central positions of the spectral lines and their order, $n$, see \autoref{fig-spectral-positions}. 
Our theoretical analysis employed a fixed (calculated) surface potential, which made it possible to determine $B_\text{He}$ by optimizing the match of the calculated \cost($t$) to the experimental trace. The found value of $B_\text{He}$, which represents the first experimental determination of the rotational constant for a molecule located on a He droplet surface, is reduced compared to $B_\text{gas}$: $B_\text{gas}$/$B_\text{He} = 1.3$. In comparison, for linear molecules, like \ce{I2}, \ce{OCS} or \ce{N2O}, localized in the interior of He droplet, $B_\text{gas}$/$B_\text{He}$ lies between 2 and 6~\cite{chatterley_rotational_2020,grebenev_rotational_2000,nauta_rotational_2001}. The smaller reduction factor, 1.3,  for \ce{Na2} reflects the fact that its interaction with the He droplet is weaker than for molecules inside He droplets~\cite{toennies_superfluid_2004, barranco_helium_2006}.

\Autoref{fig-potentials-probability} shows that $V_\text{He}$ pins \ce{Na2}(1~$^3\Sigma_{u}^+$) to the droplet surface, which indicates that the laser-induced rotational motion preferentially occurs in the tangential plane to the surface at the point where the dimer resides. Inspired by this, we calculated the alignment dynamics for a 2D rotor and compared it with the experimental results. In practice, we solved the time-dependent rotational Schr\"{o}dinger equation for \ce{Na2} modelled as a 2D quantum rotor~\cite{mirahmadi_quantum_2021}. 
Again, the rotational constant, now denoted $B_\text{2D}$ was treated as a free parameter. \Autoref{fig-traces}(c1) depicts \cost($t$) for $B_\text{2D}$ = 1.33 GHz, which produces a very close match of the positions of the maxima and minima to those in the experimental trace. Furthermore, comparison of \autoref{fig-traces}(c1) and \autoref{fig-traces}(b1) shows that \cost($t$) calculated by the 2D model and by the 3D + $V_\text{He}$ model, \autoref{eqn:2a}, are essentially indistinguishable. The same similarity is also evident for the corresponding spectra, i.e. \autoref{fig-traces}(c2) versus \autoref{fig-traces}(b2), and by the agreement between the red triangles and blue crosses in \autoref{fig-spectral-positions}. As such, the measured alignment dynamics is well-described by a 2D rotor model with a rotational constant $B_\text{2D}$, which is larger than $B_\text{He}$. In a classical picture the increase of the rotational constant can be explained as the dimer fluctuating out of the tangential plane according to $p(\Theta)$ depicted in \autoref{fig-potentials-probability}. When we average $B$ over $p(\Theta)$, we find a small increase, $\langle B\rangle / B=1.05$. This value matches the ratio $B_\text{2D} / B_\text{He}=1.05$.

Finally, to emphasize the difference between the observed alignment dynamics of the sodium dimer on the He droplet surface and that expected for gas-phase dimers, we calculated \cost($t$) for a gas of isolated \ce{Na2}(1~$^3\Sigma_{u}^+$) molecules, i.e. using \autoref{eqn:2a} without $V_\text{He}$, but otherwise with the same temperature and laser parameters. The result for $B = B_\text{He}$ is depicted in \autoref{fig-traces}(d1)-(d2). The calculated \cost($t$) qualitatively differs from the experimental trace and cannot reproduce it no matter which value of $B$ is chosen. The same mismatch holds for the positions of the peaks in the spectrum, illustrated by comparing \autoref{fig-traces}(a2) and (d2) or, equivalently, the green squares and black circles in \autoref{fig-spectral-positions}.

Returning to the two motivations for our work, we conclude that nonresonant fs laser pulses can set molecules on the surface of He nanodroplets into rotation and create a noticeable degree of alignment after the pulse is turned off.  Theoretical analysis showed that the observed nonadiabatic alignment dynamics results from the combined action of the laser-induced polarizability interaction and the surface potential. For \ce{Na2}(1~$^3\Sigma_{u}^+$) it closely resembles that of a 2D quantum rotor or, equivalently, a particle on a ring, one of the most basic quantum motions. We demonstrated that rotational coherence spectroscopy is possible and used it to determine the He-dressed rotational constant for \ce{Na2}(1~$^3\Sigma_{u}^+$). Studying other alkali dimers in either the 1~$^3\Sigma_{u}^+$ or the 1~$^1\Sigma_{g}^+$ state offers an opportunity to explore how the strength of the surface potential influences the laser-induced alignment dynamics. Preliminary studies of \ce{Na2} and \ce{K2} in the 1~$^1\Sigma_{g}^+$ state, where the dimers are less pinned to the surface than for \ce{Na2}(1~$^3\Sigma_{u}^+$), reveal alignment dynamics qualitatively different from both the results presented here and from that of isolated molecules. Other novel opportunities emerging from our work include 1) Identifying the broadening mechanisms of the rotational energy levels, causing the gradual decay of the observed rotational dynamics. 2) Time-resolved studies of transfer of rotational energy from the dimer to the He droplet. 3) Alignment in the adiabatic regime, exploring if the combination of the 0.37 K temperature and the very large polarizability anisotropies of alkali dimers~\cite{deiglmayr_calculations_2008} lead to exceptional large degrees of alignment. The work presented here is now published, see~\cite{Lorenz-2023}.

\begin{acknowledgments}
H.S. acknowledges support from The Villum Foundation through a Villum Investigator Grant No. 25886. M.L.~acknowledges support by the European Research Council (ERC) Starting Grant No.~801770 (ANGULON). F.J. and R.E.Z. acknowledge support from the Centre for Scientific Computing, Aarhus and the JKU scientific computing administration, Linz, respectively.
\end{acknowledgments}



%

\end{document}